\documentclass[conference]{IEEEtran}
\IEEEoverridecommandlockouts
\usepackage{cite}
\usepackage{amsmath,amssymb,amsfonts}
\usepackage{algorithmic}
\usepackage{graphicx}
\usepackage{textcomp}
\usepackage{xcolor}
\usepackage{lipsum} 
\usepackage{booktabs}
\usepackage{multirow}
\usepackage{caption}
\usepackage{siunitx}
\usepackage{epsfig}
\def\BibTeX{{\rm B\kern-.05em{\sc i\kern-.025em b}\kern-.08em
    T\kern-.1667em\lower.7ex\hbox{E}\kern-.125emX}}
\begin{document}

\title{A Framework for QoS of Integration Testing in Satellite Edge Clouds\\
\thanks{This work was supported in part by the National Natural Science Foundation of China under Grant 62372163, Natural Science Foundation of Chongqin under Grant ZE20210011, Key scientific and technological research and development plan of Hunan Province under Grant 2022NK2046, Yuelushan Industrial Innovation Center Cultivation Project under Grant 2023YC110130. Corresponding author: Juan Luo (juanluo@hnu.edu.cn). }
}

\author{
	\IEEEauthorblockN{Guogen Zeng, Juan Luo, Yufeng Zhang, Ying Qiao, Shuyang Teng}
	\IEEEauthorblockA{
		\textit{College of Computer Science and Electronic Engineering} \\
		\textit{Hunan University}\\
		Changsha, China \\
		\{zengguogen, juanluo, yufengzhang, qy2020, S2110Z0004\}@hnu.edu.cn
	}
}

\maketitle

\begin{abstract}
\textbf{The diversification of satellite communication services imposes varied requirements on network service quality, making  quality of service (QoS) testing for microservices running on satellites more complex. Existing testing tools have limitations, potentially offering only single-functionality testing, thus failing to meet the requirements of QoS testing for edge cloud services in mobile satellite scenarios. In this paper, we propose a framework for integrating quality of service testing in satellite edge clouds. More precisely, the framework can integrate changes in satellite network topology, create and manage satellite edge cloud cluster testing environments on heterogeneous edge devices, customize experiments for users, support deployment and scaling of various integrated testing tools, and publish and visualize test results. Our experimental results validate the framework's ability to test key service quality metrics in a satellite edge cloud cluster.}\par
\end{abstract}


\section{Introduction}
As the demand for satellite computing services continues to grow, satellites are poised to offer a wider range of service types than ever before, and different services will have varying quality of service (QoS) requirements\cite{10234306}. However, in the process of providing various services in the satellite cluster, a prevailing issue is that a large number of users making repeated service requests can often lead to network congestion and performance degradation\cite{9120643}. Additionally, devices providing software services within satellite clusters may face resource constraints or be affected by high-speed mobility, potentially resulting in QoS degradation or unresponsive services. \par
Testing the interactions of microservice applications between nodes in a satellite edge cloud can only be accomplished by running end-to-end tests or extensive integration tests, which require deploying microservice applications across the cluster. Performing end-to-end testing on satellites is exceedingly complex, time-consuming and costly. Extensive integration testing can be conducted either within or outside the satellite cluster. However, external testing has significant drawbacks, such as inefficiency, primarily because most satellite services lack public interfaces that allow access from outside the cluster\cite{9825810}. Therefore, the standard approach to extensive integration testing of microservices on satellites is to test them from within the cluster. However, some testing tools are limited, lacking key functionalities for service quality testing, or are difficult to deploy and accurately assess test content. \par

To this end, in view of satellite edge cloud service quality testing and the limitations of existing testing tools, we propose a new service quality integrated testing framework suitable for edge cloud microservice applications in satellite scenarios. The framework possesses the capability to integrate inter-satellite network topology changes, it also can create and manage satellite edge cloud cluster testing environments on heterogeneous edge devices. Furthermore, this framework can integrate and expand multiple testing tools into a unified framework, supporting customizable input parameters for testing and publishing test results. We demonstrated the feasibility and effectiveness of the testing framework.\par

\begin{figure}[t] 
	\centering
	\includegraphics[width=\columnwidth]{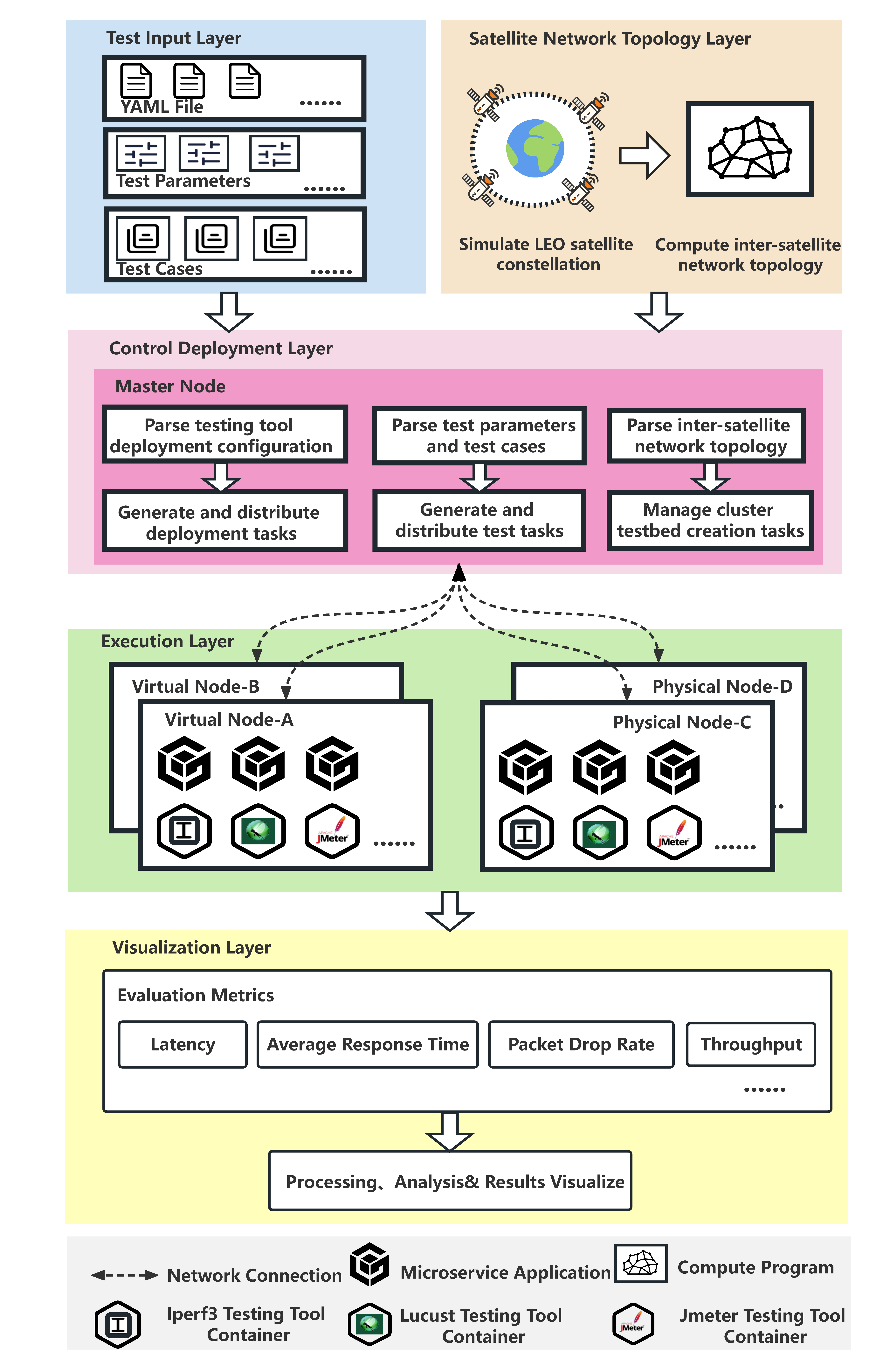} 
	\caption{Architecture of the integration testing framework}
	\label{fig:SecTest Framework }
\end{figure}

\section{TESTING FRAMEWORK}
\subsection{Description of Integration Testing Framework}
Fig. 1 provides a detailed illustration of the testing framework and  interaction process. The framework is divided into five main layers: test input layer, satellite network topology layer, control deployment layer, execution layer and visualization layer.\par
\textbf{Test Input Layer}: This layer is equipped with functionalities for user interaction, responsible for customizing test parameters and test case inputs, as well as uploading YAML configuration files for deploying testing tools.\par
\textbf{Satellite Network Topology Layer}: Within a specific timeframe set in Satellite Tool Kit (STK), the STK are utilized to simulate changes in the Low-Earth-Orbit (LEO) satellite network topology and the trajectories of satellite operations, thereby obtaining data on inter-satellite distances and communication ranges. Subsequently, we have developed a program to retrieve and process the simulated data from STK, calculating the network topology of the satellites within the defined timeframe. This data is then passed to the control deployment layer for managing the number of edge cloud clusters.\par
\textbf{Control Deployment Layer}: The test cluster generated by the testing framework follows a Master-Worker architecture. And the control deployment layer is positioned on the master node, which primarily serves the following functions: 1) It is responsible for the integrated deployment and configuration of test tools, parsing them into specific deployment tasks, and distributing them to each worker node in the execution layer. 2) It is capable of parsing test parameters and test cases, generating test tasks, and uploading them to the test tool container of the worker node for execution. 3) The control deployment layer parses the acquired satellite network topology and generates a satellite edge cloud test cluster. Additionally, it manages the scope of the satellite edge cloud cluster and performs tasks such as adding or removing nodes in the cluster. It's important to note that the generated test cluster nodes encompass both physical servers and virtual servers.\par

\textbf{Execution Layer}: Worker nodes receive tasks to create cluster test environments. The test cluster comprises virtual and physical computing nodes, with each node capable of containing multiple built test tools and microservice applications. At the execution layer, edge devices are designated as worker nodes to execute test tasks. Due to Kubernetes' potential to better integrate distributed edge devices with the cloud, we have chosen to use it to manage and execute the deployment tasks of microservice applications on the edge nodes\cite{9488701}. Finally, upon completion of the testing task, the test results are returned to the visualization layer.\par

\textbf{Visualization Layer}: Responsible for retrieving and processing test data from each running test container, this layer can publish test results on the  framework frontend page, presenting processed results intuitively to the user. To broaden the capabilities of the visualization layer, we have reserved a common interface in the visualization layer of the framework to support users' custom development. This interface is designed for acquiring and processing diverse types of test data, enabling tailored display of test results according to individual requirements. \par

\section{METHODOLOGY}
\subsection{QoS Evaluation Metrics}
The testing framework can not only be adjusted according to specific test content, but it can also customize a diverse range of metrics based on the characteristics and requirements of the integrated testing tools.\par
\textbf{Throughput}: \emph{TP} represents the throughput of successfully transmitted data packets through the communication bandwidth channel \(B\). Taking into account the possibility of inter-satellite communication failures, data packets are transmitted within reach between satellites \(x\) and \(y\). This process continues until a communication mechanism cannot be established between the satellites within a time period \(T_{connect}\). Transmission delay may occur during inter-satellite communication, where transmission delay is defined as  \(T_{trans}\)\cite{9327501}:
\begin{equation}
	T_{trans}=\dfrac{D_{trans}}{B\log_2 {(1+\dfrac{P_tG_{max}^2}{k_B\tau L(xy)})}}\label{eq 1}
\end{equation}

where \(D_{trans}\) is transmission packet size between \(x\) to \(y\), \(B\) is the channel bandwidth in Hertz, \(P_t\) is transmission power, \(G_{max}\) is the peak gain of both antennas of satellite \(x\)  in the direction of their main lobe, \(k_B\) is the Boltzmann constant, \(\tau\) is the thermal noise in Kelvin, \(L(xy)\) is path loss for an ISL between satellites \(x\) and \(y\)\cite{9327501}. Inference delay refers to the total time spent by LEO satellites from the start of target recognition to the completion of the task.\par
We define \emph{TP} as\cite{2021Intelligent}:\par
\begin{equation}
	TP = \dfrac{\sum_{i=0}^n( T_{connect}) \cdot B}{\sum_{i=0}^n (T_{connect}+T_{trans})}\label{eq 3}
\end{equation}
\par
\textbf{Packet Drop Rate}: We define the probability of data packet \(D\) experiencing data reduction during transmission from satellite \(x\) to satellite \(y\), as the percentage calculated between the total lost data packets \(D_{totalPL}\) and the total successfully transmitted data packets \(D_{totalPT}\), and is defined as\cite{2021Intelligent}, \cite{6725580}:

\begin{equation}
PDR(x,y)=\dfrac{D_{totalPL}}{D_{totalPT}}\label{eq 4}
\end{equation}

\section{EXPERIMENT}
\subsection{Experimental Setup}
To simulate the operational trajectories of the satellite cluster, we utilized the STK system toolkit to model the topological changes in the StarLink satellite network. The orbital altitude was set at 550 kilometers with an inclination angle of 45 degrees. Four orbits from the Starlink constellation were selected, each containing 10 LEO satellites, resulting in a total of 40 satellites\cite{9984697}. We randomly selected a satellite node and calculated its connectivity in the satellite network topology changes within a five-minute interval. \par

Based on the aforementioned QoS evaluation metrics, we have selected iperf3 and Locust as our testing tools. These tools will be deployed by creating containers and deploying them into the same namespace on each node.\par

\subsection{Result}
We conducted tests on the packet drop rate and throughput of 5 physical edge cloud nodes within a 5-minute timeframe under conditions of 50 Mbps, 100 Mbps, and 500 Mbps bandwidth. The master node served as the data sender, distributing data to the other Worker nodes in the cluster. The results of throughput and packet drop rate tests for the worker nodes are illustrated in Fig. 2 and Fig. 3. The experimental findings indicate that an increase in bandwidth leads to an expected increase in throughput for each edge cloud node. Simultaneously, with the ability to transmit more data within the same timeframe, the likelihood of data loss during transmission also significantly increases, resulting in a rise in the packet drop rate. Interestingly, except for P-Node1, the overall packet drop rate at 100 Mbps bandwidth is higher than that at 500 Mbps bandwidth. This may be attributed to the heavier network load during data transmission on the 100 Mbps bandwidth network.\par
\begin{figure}[!t]
	\centering
	\includegraphics[width=8cm, height=6cm]{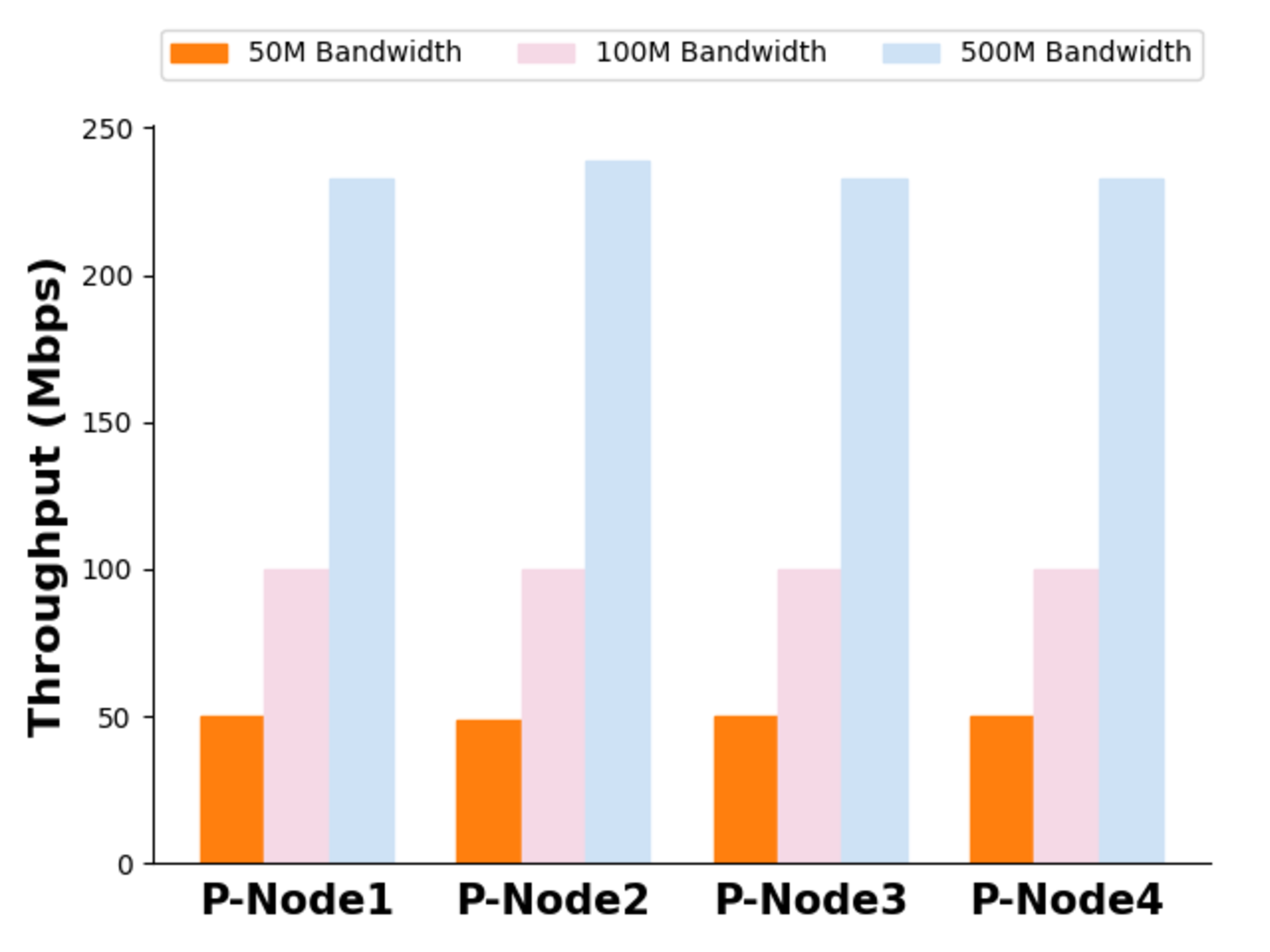}
	\caption{Throughput of different nodes at different bandwidths}
	\label{fig:Throughput}
\end{figure}
\begin{figure}[!t]
	\centering
	\includegraphics[width=9cm, height=6cm]{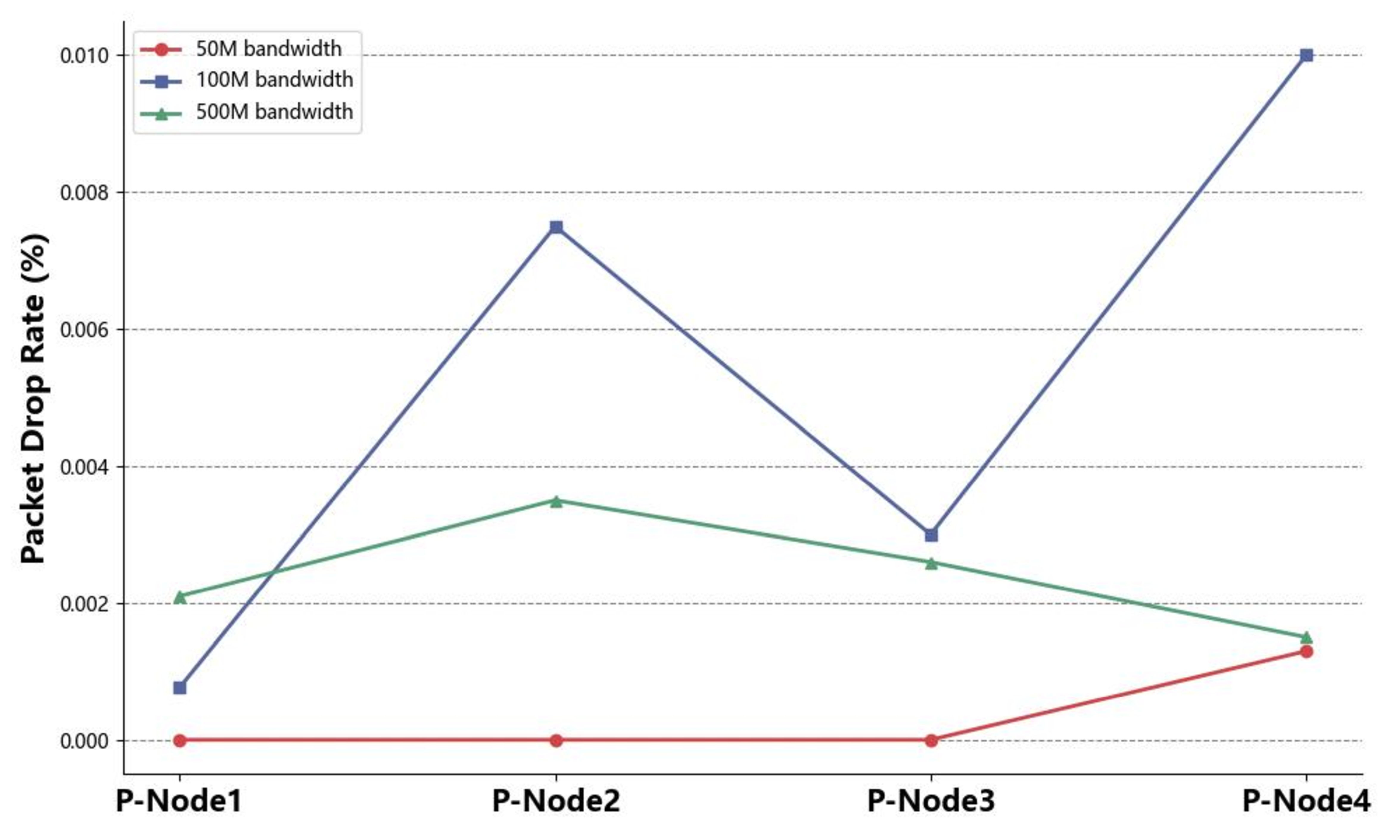}
	\caption{Packet drop rate of different nodes at different bandwidths}
	\label{fig:Packet drop rate}
\end{figure}

\section{Conclusion}
We have designed and implemented a QoS integration testing framework for satellite edge cloud environments. This framework holds significant and guiding implications for testing service quality in satellite edge cloud clusters, providing a structured and efficient method to assess the reliability and performance of these services. The framework can integrate with the changes in satellite network topology, support the management of satellite edge cloud cluster scales, integrate and extend multiple testing tools into a unified framework, and allow for custom test parameter input and publication of test results. In the future, we plan to expand the range of testing tools to provide comprehensive testing of satellite edge cloud environments from multiple perspectives.\par

\bibliographystyle{unsrt}  

\bibliography{reference.bib}  

\begin{thebibliography}{1}

\bibitem{10234306}
Chao Wang, Yiran Zhang, Qing Li, Ao~Zhou, and Shangguang Wang.
\newblock Satellite computing: A case study of cloud-native satellites.
\newblock In {\em 2023 IEEE International Conference on Edge Computing and
  Communications (EDGE)}, pages 262--270, 2023.

\bibitem{9120643}
Xinmu Wang, Hewu Liy, Wenbing Yao, Tianming Lany, and Qian Wu.
\newblock Content delivery for high-speed railway via integrated
  terrestrial-satellite networks.
\newblock In {\em 2020 IEEE Wireless Communications and Networking Conference
  (WCNC)}, pages 1--6, 2020.

\bibitem{9825810}
Christoph Reile, Mohak Chadha, Valentin Hauner, Anshul Jindal, Benjamin
  Hofmann, and Michael Gerndt.
\newblock Bunk8s: Enabling easy integration testing of microservices in
  kubernetes.
\newblock In {\em 2022 IEEE International Conference on Software Analysis,
  Evolution and Reengineering (SANER)}, pages 459--463, 2022.

\bibitem{9488701}
Yiwen Han, Shihao Shen, Xiaofei Wang, Shiqiang Wang, and Victor~C.M. Leung.
\newblock Tailored learning-based scheduling for kubernetes-oriented edge-cloud
  system.
\newblock In {\em IEEE INFOCOM 2021 - IEEE Conference on Computer
  Communications}, pages 1--10, 2021.

\bibitem{9327501}
Israel Leyva-Mayorga, Beatriz Soret, and Petar Popovski.
\newblock Inter-plane inter-satellite connectivity in dense leo constellations.
\newblock {\em IEEE Transactions on Wireless Communications}, 20(6):3430--3443,
  2021.

\bibitem{2021Intelligent}
Prohim Tam, Sa~Math, and Seokhoon Kim.
\newblock Intelligent massive traffic handling scheme in 5g bottleneck backhaul
  networks.
\newblock {\em KSII Transactions on Internet and Information Systems},
  15(3):874--890, 2021.

\bibitem{6725580}
Mohammad Esmaeilzadeh, Neda Aboutorab, and Parastoo Sadeghi.
\newblock Joint optimization of throughput and packet drop rate for delay
  sensitive applications in tdd satellite network coded systems.
\newblock {\em IEEE Transactions on Communications}, 62(2):676--690, 2014.

\bibitem{9984697}
Feng Wang, Dingde Jiang, Zhihao Wang, Jianguang Chen, and Tony Q.~S. Quek.
\newblock Seamless handover in leo based non-terrestrial networks: Service
  continuity and optimization.
\newblock {\em IEEE Transactions on Communications}, 71(2):1008--1023, 2023.

\end{thebibliography}

\end{document}